\documentclass[aps,prl,twocolumn]{revtex4}
\usepackage{amsmath,amssymb}
\usepackage[colorlinks]{hyperref}
\input epsf
\usepackage{verbatim}
\usepackage{graphicx}
\usepackage{bm}
\def\pmb#1{\setbox0=\hbox{#1}
\kern-.025em\copy0\kern-\wd0 \kern-.05em\copy0\kern-\wd0
\kern-.025em\raise.0433em\box0}

\newcommand{\beq}{\begin{equation}}
\newcommand{\eeq}{\end{equation}}
\newcommand{\ba}{\begin{eqnarray}}
\newcommand{\ea}{\end{eqnarray}}

\input epsf

\usepackage[english]{babel}
\begin{document}

\title[]{On form invariance of the Kirchhoff-Love plate equation
}
\author{L. Pomot$^{1,2}$, S. Bourgeois$^1$, C. Payan$^1$, M. Remilieux$^{3}$, S. Guenneau$^2$}
\affiliation{$^1$ Aix-Marseille Univ, CNRS, Centrale Marseille, LMA, Marseille, France}
\affiliation{$^2$ Aix-Marseille Univ, CNRS, Centrale Marseille, Institut Fresnel, Marseille, France}
\affiliation{$^3$ Geophysics Group (EES-17), Los Alamos National Laboratory, Los Alamos, New Mexico 87545, USA}
\begin{abstract}
 
We propose an alternative approach called backward transformation for the design of platonic cloaks, without resorting to in-plane body forces and pre-stresses, which can lead to unphysical features. It is shown that the Kirchhoff-Love equation for an anisotropic heterogeneous plate is actually form invariant for a class of transformations with a vanishing Hessian. This formalism is detailed and numerically validated with three-dimensional simulations in the time domain. Results published in the literature fall in the class of transformations proposed here.


\end{abstract}
\maketitle
    
There has been a growing interest over the past 10 years in the analysis of elastic waves in thin plates in the metamaterial community with the theoretical proposal \cite{farhat09,farhat09PRL}, and its subsequent experimental validation \cite{stenger12}, of a broadband cylindrical cloak for flexural waves. A square lattice cloak based on an improved transformed plate model \cite{colquitt14} has been also experimentally validated for flexural waves \cite{misseroni16}. Interestingly, this latter cloak is reminiscent of a mechanical lattice cloak designed by a heuristic transform approach, which has been experimentally validated in an elastostatic case \cite{buckmann15}.

One of the attractions of metamaterial plates (or platonics) \cite{mcphedran09} for the photonics community is that analysis of transverse electromagnetic waves in metamaterials can be sometimes translated to surface elastic waves in structured plates, within a thin plate approximation. The theory describing the flexural motion of thin homogeneous plates is well established and can be found in classical books, such as \cite{timoshenko59}. There are some subtleties in the mathematical analysis, and numerical models, of the scattering of flexural waves owing to the fourth-order derivatives involved in the governing equation, versus the usual second-order derivatives for the wave equation of optics; even the waves propagating in a perfect plate are unlike those of the acoustic and electromagnetic wave equations as they are not dispersionless. Thus numerical computations need to take into account additional limit conditions compared with second order wave equations. Moreover, when a plate is anisotropic the governing equation and limit conditions become more involved and earlier computations made for the cylindrical platonic cloak in \cite{farhat09,farhat09PRL} have been revisited \cite{brun14,climente16}. Fortunately, all computations performed thus far support the first experimental proof of platonic cloaking by Wegener's group in \cite{stenger12}. However, when one moves into the area of volume elastic waves, governed by the full Navier equations, it is no longer possible to reduce the analysis to a single scalar partial differential equation (PDE) as in the case of flexural waves, as shear and pressure waves do couple at boundaries. A key issue is that the Navier equations are generally not form invariant \cite{milton06}, or lead to a minor-symmetry breaking in the elasticity tensor \cite{brun09}, and thus the experimental validation of an elastodynamic cloak remains elusive thus far.

Within the simplified framework of the Kirchhoff-Love plate theory \cite{timoshenko59} bending moments and transverse shear forces are taken into account via a fourth-order PDE for the out-of-plane plate displacement field. This plate theory is a natural extension of the Helmholtz equation to a generic model for flexural wave propagation through any spatially varying thin elastic medium. It offers a very convenient mathematical model for any physicist wishing to study flexural waves in plates using earlier knowledge in photonics or phononics. However, while the Helmholtz equation can, with appropriate notational and linguistic changes, hold for acoustic, electromagnetic, water or out-of-plane elastic waves and so encompasses many possible  applications, the Kirchhoff-Love (KL) plate theory is dedicated to the analysis of flexural waves. Besides, as already noted, the analogy with models in photonics partially breaks down for anisotropic plates, and this is not surprising as flexural, also called bending, waves are fundamentally different in character from compressional acoustic or electromagnetic waves : they model the lowest frequency waveguide mode for elastic waves and therefore are dispersive. For these low frequencies, equivalently thin plates, the phase and group velocities scale with square root of the frequency and this is reflected in the KL equation which is fourth order in space and second order in time. In the frequency domain and for an isotropic homogeneous plate the KL equation reads as
\begin{equation}
D_0 \Delta^2 w(x_1, x_2) - \rho H \omega ^2 w(x_1,x_2) = 0
\label{Kirchhoff}
\end{equation}
with $w(x_1,x_2)$ the out-of-plane displacement, $H$ the plate thickness, $\rho$ its density and $D_0$ its rigidity. In 2009, Farhat \textit{et al.} \cite{farhat09PRL} used the theory of acoustic cloaking developped by Norris\cite{NorrisAcoustic} to deduce a transformed biharmonic equation, which can be interpreted in terms of an anisotropic rigidity of the form
\begin{equation}
\underline{\underline{D}} = D_0 F F^T F F^T |\textnormal{det}F|^{-1}
\end{equation}
with $F$ the gradient of the considered transformation. However this approach does not take into account the full complexity of flexural waves in anisotropic media described by a fourth order rigidity tensor. In 2014, Colquitt \textit{et al.}\cite{colquitt14} corrected the transformed equation by considering an appropriate anisotropic rigidity and using again the theory developed in \cite{NorrisAcoustic}. Especially the lemma (2.1) therein, was applied twice to account for the fourth order derivatives in (\ref{Kirchhoff}). In \cite{colquitt14}, the Kirchoff Love equation (\ref{Kirchhoff}) is mapped onto a von-Karman equation:
\begin{equation}
\begin{array}{l}
D_{ijkl}w_{,ijkl} + 2D_{ijkl,i}w_{,jkl} + (D_{ijkl,ij}  N_{kl})w_{,kl}\\
+ S_{l}w_{,l} = h\rho\omega^2 w
\; , \; i,j,k,l=1,2
\end{array}
\label{colquitt}
\end{equation}
where we use Einstein's summation convention, the subscript $,i...l$ denoting partial derivatives with respect to space variables $x_i$... $x_l$ and
\begin{equation}
\left\{
\begin{array}{lll}
D_{ikjl}&=&D_0 F_{ip}F_{jp}F_{km}F_{lm}  J^{-1} \\
N_{kl}&=&D_0{(JG_{kl}G_{ij,i}-JG_{jl}G_{ik,i})}_{,j}\\
&-&D_0G_{jk}{(JG_{il,i})}_{,j} \\
S_{l}&=&D_0{[G_{jk}{(JG_{il,i})}_{,j}]}_{,k} \\
\end{array} \right.
\end{equation}
with $J = \textnormal{det}F $, $G_{ij} = J^{-1} F_{ip}F_{jp}$ and $h=H|\textnormal{det}F|^{-1}$. Despite the fact that this result is more suited than the transformed equation in \cite{farhat09} to describe the full complex behaviour of flexural wave, we point out that taking a geometrical transformation equal to the identity (i.e. $F=Id$) does not give the expected rigidity of an homogeneous anisotropic plate. In fact, if the quantities $N$ and $S$ are indeed equal to the null tensor $0$, the transformed rigidity tensor does not correspond to an homogeneous anisotropic rigidity. Indeed, by using Voigt's notation we can write the transformed rigidity as:
\begin{equation}
\underline{\underline{D}}^{Id} = \begin{pmatrix}
D_0 &  D_0 & 0 \\
 D_0 & D_0 & 0 \\
0 & 0 & 0
\end{pmatrix} 
\end{equation}
which is different from the original rigidity tensor. Furthermore the obtained rigidity tensor is not positive definite, leading to zero energy deformation modes. It seems to us that there are several solutions to the equation $(D_{ijkl}w_{,kl})_{,ij} = D_0 \Delta (\Delta w)$ for $D_{ijkl}$ and that applying twice the Lemma 2.1. proposed by Norris in \cite{NorrisAcoustic} breaks the initial structure of the equation, leading to the solution proposed in \cite{colquitt14}. 

In this Letter, we would like to propose a different approach called backward (or pullback) transformation, with a one-to-one correspondence between an anisotropic heterogeneous Kirchhoff-Love equation, and an anisotropic heterogeneous von-Karman equation. Furthermore, we shall see that the Kirchhoff-Love equation is actually form invariant for certain class of transformations with a vanishing Hessian.

\begin{figure}[h!]
\includegraphics[width=9cm]{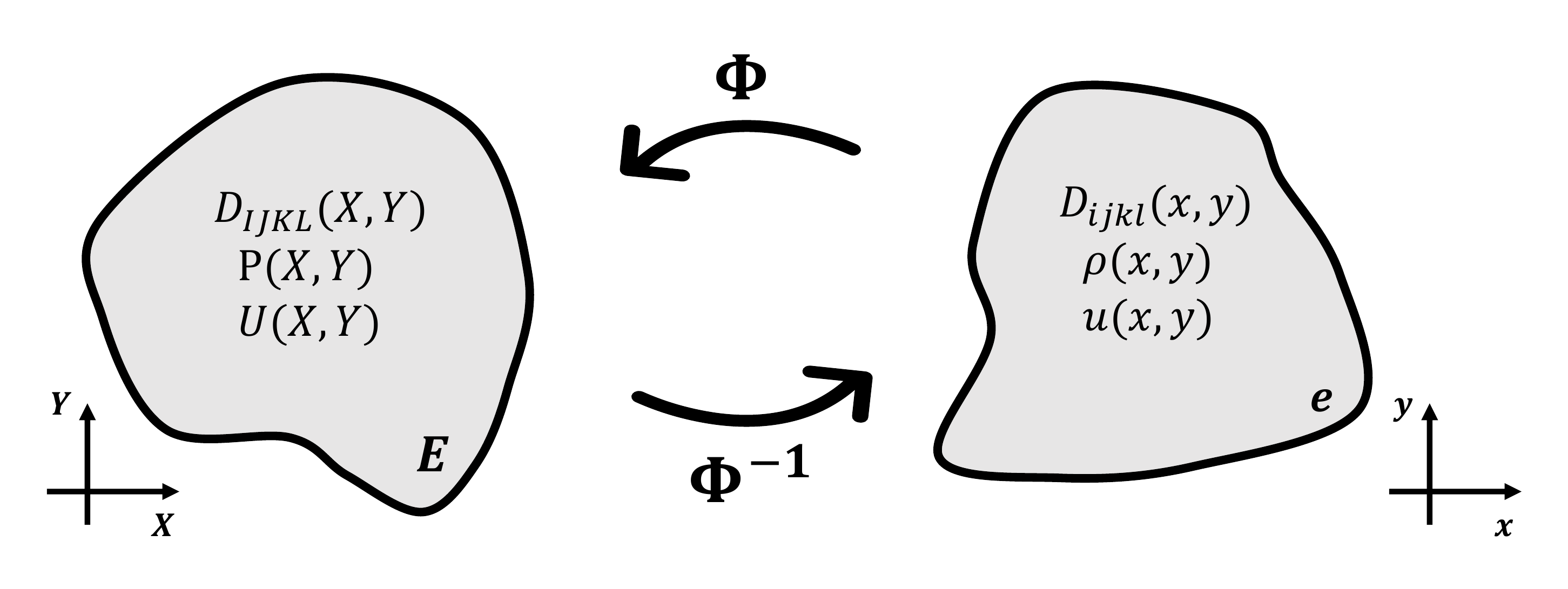}
\caption{Schematic representation of a geometrical transform that maps an anisotropic inhomogeneous propagation medium onto an other anisotropic inhomogeneous propagation medium}
\label{fig1}
\end{figure}

Notations used in this Letter are really similar to those used in \cite{Norris}  where two configurations are considered, one for the original space $E$ and one for the transformed space $e$, also called the physical space. The geometrical transformation from $e $ to $E$  is denoted by $\Phi$, upper and lower case are used to distinguish between the domains. We note $F = \nabla_x X$ the jacobian of the transformation and $J = \text{det} \Phi$ the determinant of the jacobian. $U$ and $u$ describe the infinitesimal displacements in both domains , and likewise for $D_{IJKL}$ and $D_{ijkl}$ the rigidities or $P$ and $\rho$ the mass densities. All the notations are summarized in figure \ref{fig1}. 
To avoid any loss of generality we consider first a transformation between two anisotropic inhomogeneous media where the elements of the rigidity tensors satisfy the full symmetries.  In a second part we will see a concrete example of such a transformation. One of the main ideas in this Letter is to perform a backward transformation, in the similar way to researchers introduce pull-back transforms in computational electromagnetism \cite{nicolet94}: our transformation $\Phi$ goes from the transformed space to the original space. This reflects the fact that the transformed space is generally more complex than the homogeneous space. We believe that to capture the full complexity of flexural waves the most complex equation of motion between the two spaces should be considered. By doing this we ensure the reciprocity of the transformation. \\ 
The total energy density in the transformed space is 
\begin{equation}
\mathcal{E} = \mathcal{W} + \mathcal{T} \
\label{nrj}
%
\end{equation}
where
\begin{equation}
 \mathcal{W} = \int_e \frac{1}{2} u_{,ij} D_{ijkl}(\textbf{x}) u_{,kl}\textnormal{dv}  \ \textnormal{and} \ \mathcal{T} = \int_e \frac{1}{2} \rho(\textbf{x})h \dot{u}^2 \textnormal{dv} 
\label{nrjvarform} 
\end{equation}
\\
The derivative rules lead to :
\begin{equation}
u_{,ij}=F_{Ii} F_{Jj}  U_{,IJ}+ \Gamma_{Iij} U_{,I}
\end{equation}
with $\Gamma_{Iij}=\frac{\partial^2 X_I}{\partial x_i \partial x_j}$ , the Hessian matrix. Hence the energy density in the initial space reads $\mathcal{E}_0  = \mathcal{W}_0 + \mathcal{T}_0 $ with 
\begin{equation}
\begin{split}
 \mathcal{W}_0 =& \int_E \frac{1}{2} \left( U_{,IJ}F_{Ii}F_{Jj}+  U_{,I} \Gamma_{Iij} \right) D_{ijkl}(\textbf{X}) \\ 
 & \quad \left( U_{,KL}F_{Kk}F_{Ll}+ U_{,K} \Gamma_{Kkl}  \right) J^{-1}\textnormal{dV}
 \end{split}
 \label{nrjvarformtransfor1}
\end{equation}
and 
\begin{equation}
   \mathcal{T}_0= \int_E \frac{1}{2} P (\textbf{X}) \dot{U}^2 J^{-1} \textnormal{dV}
   \label{nrjvarformtransfor2}
\end{equation}
The equations of motion can be derived from the stationarity conditions of the Lagrangian density $\mathcal{L}_0=\mathcal{W}_0-\mathcal{T}_0$. One obtains :
\begin{equation}
\begin{split}
\delta \mathcal{W}_0 &=  \int_E \left( \delta U_{,IJ}F_{Ii}F_{Jj}+  \delta U_{,I} \Gamma_{Iij} \right) D_{ijkl} (\textbf{X}) \\ 
& \qquad \left( U_{,KL}F_{Kk}F_{Ll}+ U_{,K} \Gamma_{Kkl}  \right) J^{-1}\textnormal{dV} \\
\end{split}
\end{equation}
Integrating by parts, we have (omitting the boundary terms) : 
\begin{equation}
\begin{split}
\delta \mathcal{W}_0 &= \int_E \delta U \left(J^{-1} D_{ijkl} F_{Ii} F_{Jj} F_{Kk} F_{Ll} U_{,KL}\right) _{,IJ} \textnormal{dV} \\
&+ \int_E \delta U \left(J^{-1} D_{ijkl} F_{Ii} F_{Jj}  \Gamma_{Kkl} U_{,K}\right) _{,IJ}  \textnormal{dV} \\ 
&- \int_E \delta U \left(J^{-1} D_{ijkl} \Gamma_{Iij}  F_{Kk} F_{Ll} U_{,KL}\right) _{,I}  \textnormal{dV} \\ 
&- \int_E \delta U \left(J^{-1} D_{ijkl} \Gamma_{Iij} \Gamma_{Kkl}  U_{,K}\right) _{,I}  \textnormal{dV} \\ 
\end{split}
\end{equation}
Using the symmetries of $ D_{ijkl}$, one gets
\begin{equation}
\delta \mathcal{W}_0 = \int_E \delta U \left(- M_{IJ,IJ} -N_{IJ}U_{,IJ}+S_{I}U_{,I} \right) \textnormal{dV} \\
\end{equation}
with
\begin{equation}
\left\{
\begin{array}{l}
M_{IJ}=-D_{IJKL} U_{,KL} \\
D_{IJKL} = J^{-1} D_{ijkl} F_{Ii} F_{Jj} F_{Kk} F_{Ll} \\
N_{IJ}=-\left(  J^{-1} D_{ijkl} \left( 2 F_{Ki} F_{Jj} \Gamma_{Ikl}- \Gamma_{Kij}F_{Ik}F_{Jl}\right)\right)_{,K}\\
\phantom{N_{IJ}=-}  +J^{-1} D_{ijkl} \Gamma_{Jij} \Gamma_{Ikl} \\
S_I=\left(J^{-1} D_{ijkl} F_{Ji} F_{Kj}  \Gamma_{Ikl}\right)_{,JK}\\
\phantom{S_I=-}\left(J^{-1} D_{ijkl} \Gamma_{Jij} \Gamma_{Ikl}  \right)_{,J}
\end{array}
\right.
\label{eq:transformedParameters}
\end{equation}
The equations of motion in the original space is then given by 
\begin{equation}
(M_{IJ})_{,IJ}+N_{IJ}U_{,IJ}-S_{I}U_{,I}  = J^{-1} P h  \ddot{U}.
\end{equation}
Our first observation is that the transformed rigidity tensor verifies the major and minor symmetries as long as the original tensor does. This will always be the case in practice as one of the tensors will correspond to the homogeneous rigidity tensor. However, unlike the result derived in \cite{colquitt14}, the transformed equation does not verify in general the equilibrium equation $N_{IJ,J}+S_I = 0$. To overcome this obstacle we observe that this equation of motion can be greatly simplified for geometrical transformations that verify $\Gamma = 0$ , i.e. transformations with a uniform gradient. In this trivial case we have $N = 0$ and $S = 0$, leading to an equation of motion identical to the one in the original space, the equation is then form invariant.  We shall see now illustrative examples of such a transformation. 

\begin{figure}[h!]
\includegraphics[width=9.5cm]{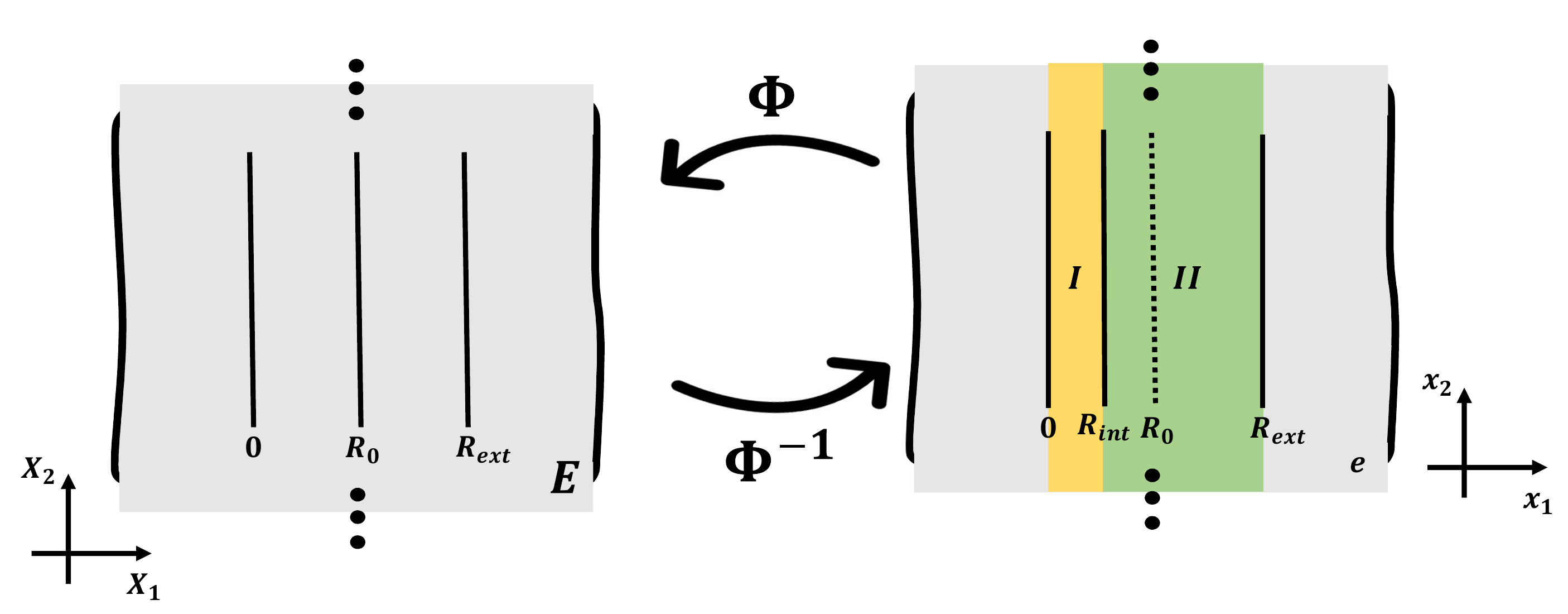}
\caption{Geometrical transformation introduced in \cite{diamondCloak}. A slab is compressed and the other is extended simultaneously, creating two anisotropic homogeneous materials. Each color corresponds to a different material.}
\label{fig:transfo1D}
\end{figure}

We start with the simple case of a 1D transformation as introduced by \cite{gralak16}. This transformation compresses a slab of size $R_0$ onto a slab of thickness $R_{int}$ and extend a slab of thickness $R_{ext}- R_0$ onto a slab of thickness $R_{ext}-R_{int}$ (see figure \ref{fig:transfo1D}), creating two homogeneous anisotropic materials. In this case we consider an infinite plate in the dimension not affected by the transformation.\\
Let us treat the case of the transformation for the material $I$. The associated geometrical transformation is : 
\begin{equation}
\Phi_{I}^{-1} : \left( \begin{array}{c} X_1 \\ X_2 \end{array} \right) \rightarrow  \left( \begin{array}{c} x_1 = \frac{R_{int}}{R_0} X_1 \\ x_2 = X_2 \end{array} \right)
\end{equation}
Looking at Hessian $\Gamma$ we have : 
\begin{equation}
 \Gamma_{Iij} = \frac{\partial^2 X_I}{\partial x_i \partial x_j} = 0 \ \ \  \forall I,i,j =1,2.
\end{equation}
Thus the only quantity to determine is the rigidity tensor $D_{ijkl}$, using equation \ref{eq:transformedParameters} and the Voigt's convention we have:  
\begin{equation}
D^{I} = \begin{pmatrix}
D_0 \frac{R_{int}^3}{R_0^3} & \nu D_0  \frac{R_{int}}{R_0} & 0 \\
\nu D_0  \frac{R_{int}}{R_0} & D_0 \frac{R_0}{R_{int}}& 0 \\
0 & 0 & \frac{1 - \nu}{2} D_0  \frac{R_{int}}{R_0}
\end{pmatrix}
\end{equation}
where $\nu$ is the Poisson ratio.

If we consider the case $R_{int} = R_0$, i.e. a transformation equal to the identity, the rigidity tensor becomes: 
\begin{equation}
D^{I} = \begin{pmatrix}
D_0 & \nu D_0 & 0 \\
\nu D_0 & D_0 & 0 \\
0 & 0 & \frac{1 - \nu}{2} D_0
\end{pmatrix} = D_{hom}.
\label{eq:Dhom}
\end{equation}

\begin{figure}[h!]
\includegraphics[width=9.5cm]{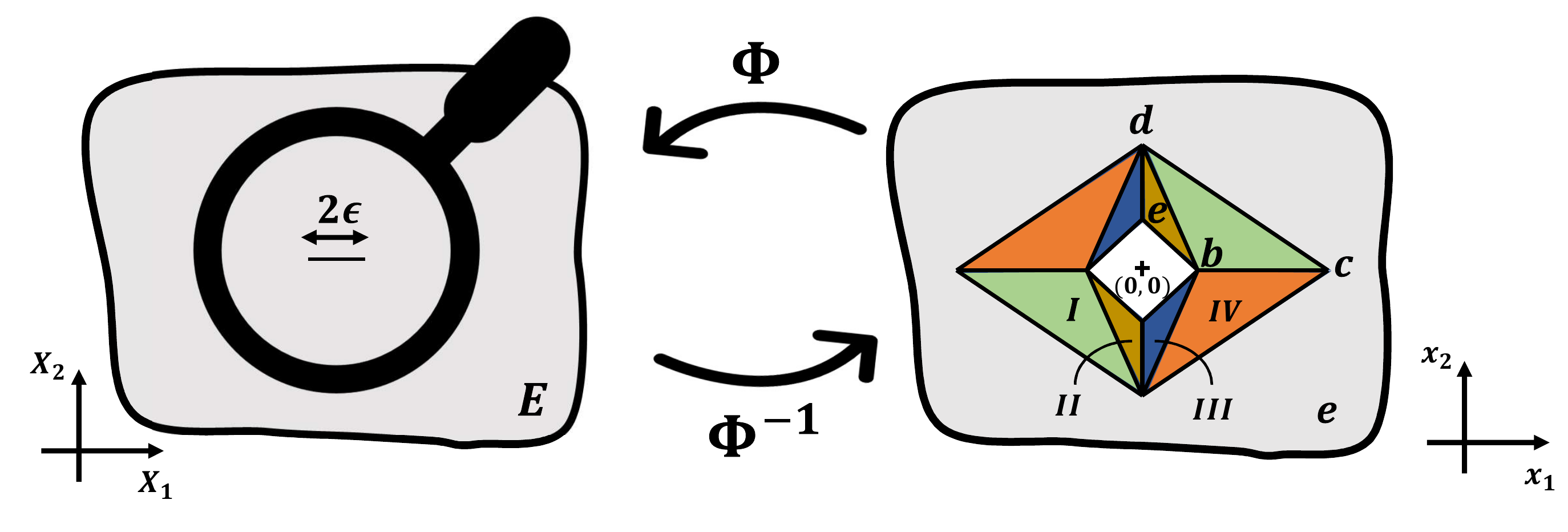}
\caption{Geometrical transformation introduced in \cite{diamondCloak}. A small segment of size $2\epsilon$ is sent onto a quadrilateral, creating four anisotropic homogeneous materials. Each color corresponds to a different material.}
\label{fig2}
\end{figure}

\begin{figure}[h!]
\includegraphics[width=9.5cm]{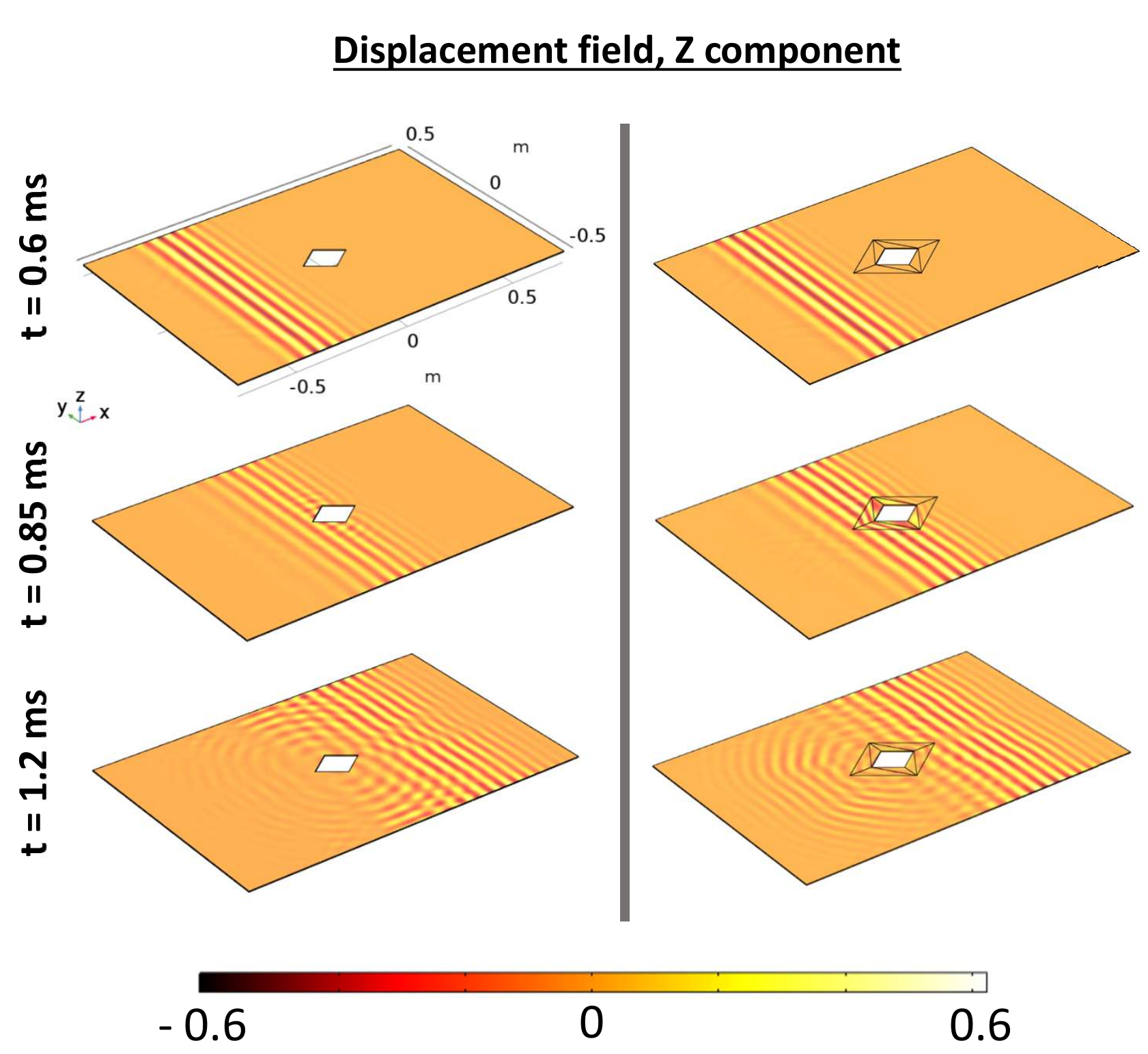}
\caption{Numerical results showing the propagation of a plane wave in a plate through anisotropic materials obtained via the geometrical transformation introduced in figure \ref{fig2}. We display here the vertical component of the displacements. Left panel : homogeneous plate with a hole. Right panel : plate with the same hole cloaked.
}
\label{fig:resultNumerique}
\end{figure}

Another example of such an invariant transformation is introduced by Li \textit{et al.}\cite{diamondCloak}. This transformation sends a small notch of size $2\epsilon$ to a diamond cloak made of 4 different anisotropic homogeneous materials (see figure \ref{fig2}).  \\
As an example we will treat the case of the transformation for the material $II$. The associated geometrical transformation is (considering $\epsilon$ small in comparison to all other characteristic length): 
\begin{equation}
\Phi_{II}^{-1} : \left( \begin{array}{c} X_1 \\ X_2 \end{array} \right) \rightarrow  \left( \begin{array}{c} x_1 = \frac{b}{\epsilon} X_1 \\ x_2 = \frac{d-e}{d} X_2 - \frac{e}{\epsilon} X_1 + e  \end{array} \right)
\end{equation}
Looking at the Hessian $\Gamma$ we note that : 
\begin{equation}
 \Gamma_{Iij} = \frac{\partial^2 X_I}{\partial x_i \partial x_j} = 0 \ \ \  \forall I,i,j=1,2.
\end{equation}
Thus the only quantity to be determined is the rigidity tensor $D_{ijkl}$. Using equation \ref{eq:transformedParameters} and the Voigt convention we find:  
\begin{equation}
\left\{
\begin{array}{l}
 D^{II}_{11} = \frac{b^3 d D_0}{\epsilon^3 (d - e)}\\
 D^{II}_{12} =  D^{II}_{21} = \frac{b D_0 (d^2 e^2 +   \nu \epsilon^2 (d-e)^2)}{\epsilon^3 d (d - e)} \\
 D^{II}_{13} = D^{II}_{31} =  - \frac{b^2 d D_0 e}{\epsilon^3 (d - e)} \\
 D^{II}_{22} = \frac{D_0 (\epsilon^2 d^2 - 2 \epsilon^2 d e + \epsilon^2 e^2 + d^2 e^2)^2}{\epsilon^3 b d^3 (d - e)} \\
D^{II}_{23} = D^{II}_{32} =  - \frac{D_0 e (\epsilon^2 d^2 - 2 \epsilon^2 d e + \epsilon^2 e^2 + d^2 e^2)}{\epsilon^3 d (d - e)} \\ 
D^{II}_{33} = - \frac{b D_0 (-\epsilon^2 d^2 + 2 \epsilon^2 d e - \epsilon^2 e^2 - 2 d^2 e^2 + \epsilon^2 d^2 \nu - 2 \epsilon^2 d e \nu + \epsilon^2 e^2 \nu)}{2 \epsilon^3 d (d - e)}
\end{array}
\right.
\end{equation}
Once more, if we consider the case $b = \epsilon$ and $e = 0$, i.e. a transformation equal to the identity, the rigidity tensor corresponds as it should to the original tensor associated with a homogeneous anisotropic plate as in equation \ref{eq:Dhom}. We implement these coefficients in the solid mechanics module of the finite element package comsol multiphysics 5.3. We consider a full elastic 3D medium and performed a temporal simulation. The medium of propagation in the solid mechanics module is defined but the elasticity tensor $C_{ijkl}$ that is not provided with our theory. However, using the definition of the bending stiffness $D_{\alpha \beta} = \int_{-h}^h x_3^2 C_{\alpha \beta} dx_3$ (where the voigt notation is used) we can express most of the components of the elasticity tensor. The remaining components are chosen in order to minimize the coupling between the vertical and the horizontal displacements ($C_{13} = C_{23} = 0$). This is probably not the optimal approximation and the following numerical results (fig \ref{fig:resultNumerique}) may be improved with a better conversion from the rigidity to the elasticity tensor. However, we stress that these results have been obtained from full 3D temporal simulations, and similar results were not reported before to the best of our knowledge. The signal sent is a Ricker pulse with a center frequency $f_c = 5000$ Hz.  
It is worth noticing that the wavefront is well distorted into the cloak, and exits as a quasi perfect plane wave. However, it is also noticeable that a visible amount of the energy is scattered backward. This point remains under study and can be attributed to the approximation $C_{13}=C_{23}=0$. However, using a full 3D numerical model associated with time domain simulations, our results demonstrate the ultra broad band nature of the designed cloak while former studies rather focus on monochromatic signals.\\

\begin{figure}[h!]
\includegraphics[width=9cm]{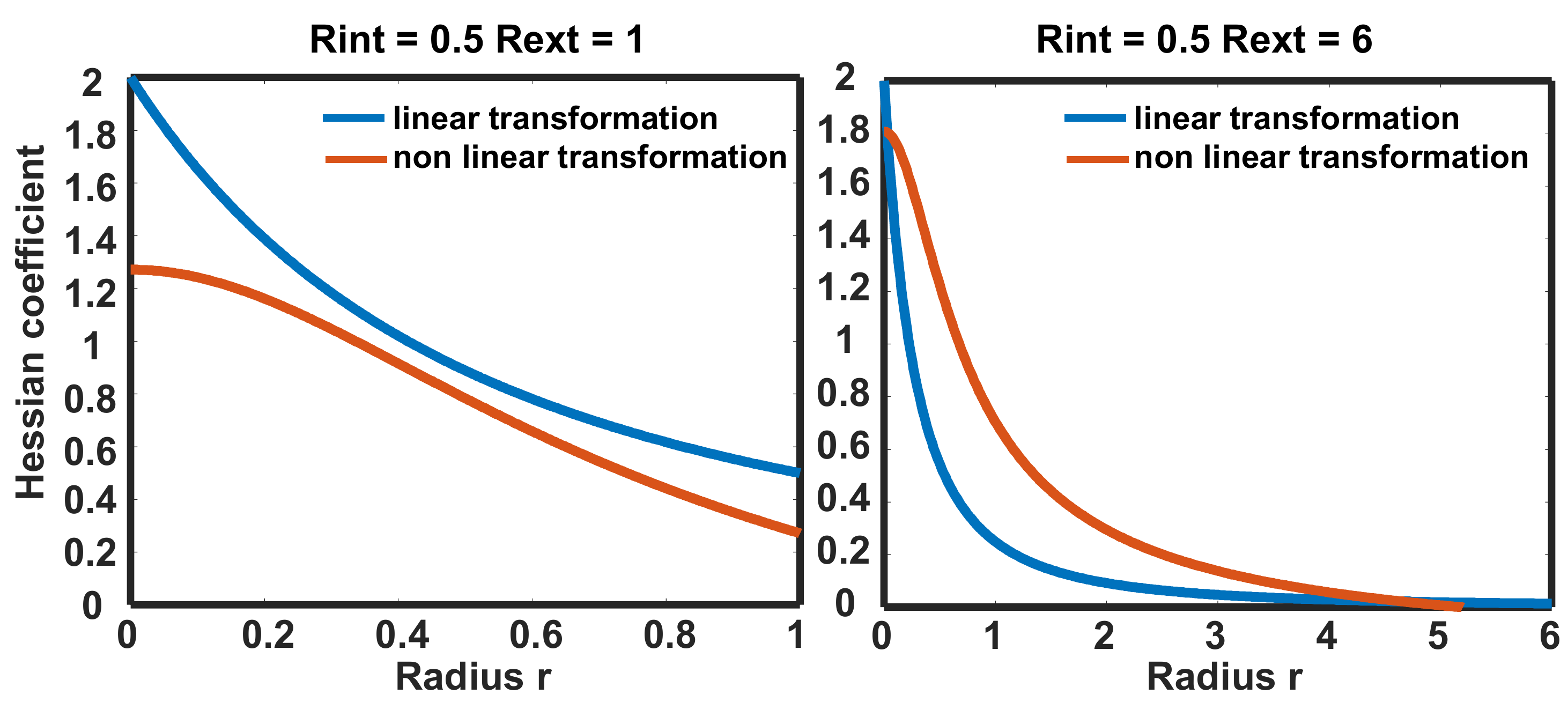}
\caption{Comparison of the variations of the Hessian coefficient $\Gamma_{Rrr}$ of linear \cite{farhat09,farhat09PRL,stenger12,brun14} and non linear\cite{nonLinearTransfo} transformations for a ratio $\frac{R_{int}}{R_{ext}}=2$ (left panel) and close to zero (right panel). }
\label{fig:coeffHessian}
\end{figure}

Finally, we would like to revisit the case of a cylindrical platonic cloak and show how our approach is in accordance with the literature. In \cite{brun14}, a linear cylindrical transformation is used to obtain numerical results. The Hessian of this linear transformation is governed by the term $ \frac{\partial}{\partial R} (\frac{R}{r}) $ with $r = \frac{R_{ext} - R_{int}}{R_{ext}} R + R_{int}$ and is not equal to zero. However, we observe that the value of the Hessian coefficient gets closer to 0 when $R_{ext} \gg R_{int}$, which is an assumption made in \cite{brun14}.  Thus, by assuming $R_{ext} \gg R_ {int}$ the authors increased the efficiency of their cloak by reducing the values of the Hessian coefficients of the transformation they considered. \\
In \cite{nonLinearTransfo}, the authors show how one can improve the efficiency of a cylindrical platonic cloak by considering a non linear transformation. The authors consider the KL equation for a plate of varying thickness, which is reminiscent of some directional cloak designed in a plate structured with a forest of rods of varying heights \cite{colombi15}. 
Provided that $R_{ext} \textcolor{green}{\gtrapprox} R_{int}$ we point out that the Hessian coefficients of the non linear transformation tend to be smaller than the coefficients of the linear transformation provided that $R_{ext}$ is not very large in comparison to $R_{int}$. An interesting result is that the coefficients of the non linear transformation tend to be larger than those of the linear transformation when $R_{ext} \gg R_{int}$.  The comparison of the coefficients is shown on figure \ref{fig:coeffHessian} for two ratios between $R_{ext}$ and $R_{int}$. It appears that depending on the ratio between these parameters, a linear transformation can be actually more effective than a non linear one.

In conclusion, making use of a backward transformation in the tensorial Kirchhoff-Love equation, we manage to capture the full complexity of the behavior of the flexural waves. Doing so preserve the reciprocity of the transformation when applied to an anisotropic plate. This is unlike what has been done in earlier papers \cite{farhat09,farhat09PRL,colquitt14,brun14}, where a forward transform was applied to a Kichhoff-Love equation in a homogeneous isotropic plate. The transformed plate model of \cite{colquitt14} improves those from \cite{farhat09,farhat09PRL} thanks to identification of some in-plane body forces and pre-stress in an anisotropic heterogeneous von-Karmann equation. However some vanishing eigenvalues of the rigidity tensor remain hard to interpret physically. Furthermore we show that for some class of transformations with a vanishing Hessian $\Gamma$, the  Kirchoff Love equation can be considered form invariant. The simplified plate equation derived in \cite{brun14} under the assumption $R_{ext} \gg R_{int}$  is naturally retrieved in our framework, as well as the improvement observed in \cite{nonLinearTransfo} between a linear and a non linear transformation. In this Letter we focus on a trivial class of transformations that verify the form invariance of the Kirchoff-Love equation. However, any transformation fulfilling the condition $N=0$ and $S=0$ would make the Kirchhoff-Love equation form invariant. 

\end{document}